\documentclass[3p,times,twocolumn]{elsarticle}

\usepackage{ecrc}

\volume{00}

\firstpage{1}

\journalname{Nuclear Physics B Proceedings Supplement}

\runauth{M. Kakizaki}

\jid{nuphbp}

\jnltitlelogo{Nuclear Physics B Proceedings Supplement}

\usepackage{amssymb}

\usepackage[figuresright]{rotating}

\begin{document}

\begin{frontmatter}



\dochead{}

\title{Higgs Phenomenology of the Supersymmetric Grand Unification with the Hosotani Mechanism
\tnoteref{label1}
}
\tnotetext[label1]{Talk presented at
    the 37th International Conference on High Energy Physics (ICHEP 2014),
    Valencia, Spain, 2-9 July 2014.  This talk is based on the work
    in collaboration with Shinya Kanemura, Hiroyuki Taniguchi and
    Toshifumi Yamashita \cite{SGGHU}.}


\author{Mitsuru Kakizaki}
\ead{kakizaki@sci.u-toyama.ac.jp}

\address{Department of Physics, University of Toyama, Toyama 930-8555, Japan}

\begin{abstract}

  The supersymmetric $SU(5)$ grand unified theory with the gauge
  symmetry broken by the Hosotani mechanism naturally solves the mass
  hierarchy problem between the colored Higgs triplet and the
  electroweak Higgs doublet, and predicts the existence of adjoint
  chiral superfields with masses of the order of the supersymmetry
  breaking scale as a byproduct.  In addition to the two $SU(2)_L$
  Higgs doublets of the minimal supersymmetric standard model, the
  Higgs sector is extended by an $SU(2)_L$ triplet chiral
  supermultiplet with hypercharge zero and a neutral singlet one.
  Such new triplet and singlet chiral supermultiplets deviate the
  standard model-like Higgs boson couplings and the additional Higgs
  boson masses from their Standard Model predictions.  We show that
  this model can be distinguished from other new physics models using
  by precisely measuring such Higgs couplings and masses, and that our
  model is a good example of grand unification testable at the
  luminosity up-graded Large Hadron Collider and future
  electron-positron colliders.

\end{abstract}


\end{frontmatter}


\section{Introduction}

One of the most significant development in the past decades
is the discovery of a new particle with mass of approximately 125 GeV,
which was announced by
the ATLAS and CMS collaborations at the CERN Large Hadron Collider 
(LHC) in 2012 \cite{LHC}. 
The spin and CP properties as well as the
couplings have been analyzed, and it has been shown that 
the nature of the discovered particle is consistent with the Standard Model (SM)
Higgs boson.  
Therefore, the SM is established as a low energy
effective theory that consistently explains phenomena below the 
TeV scale.

However, the SM have problems that should be resolved in a
more fundamental theory. 
Since the SM Higgs boson is an elementary scalar,
an unnatural cancellation between its bare
mass squared and quadratically divergent contributions from radiative
corrections is required.  
in order to keep the Higgs boson mass to the weak scale.
The reason for the fact that the electric charges of the SM
particles are fractionally
quantized is not explained.

It is intriguing that that the above-mentioned problems
can be solved by introducing the concepts of supersymmetry
(SUSY) and grand unification \cite{GUT,SUSY-GUT}.
In supersymmetric extensions of the SM, 
the loop contributions from SM particles are canceled with those
from superpartners, and 
the problem of the quadratic divergence in the Higgs boson mass squared
is avoided.
In Grand Unified Theories (GUTs), 
the SM gauge groups is embedded into a larger gauge group.
Simultaneously, the SM fermions are also embedded into larger representations.  
If the GUT gauge group is (semi-)simple, 
quantization of the electric charges of the
SM particles is automatically realised.  
Therefore, models of SUSY GUTs 
are excellent candidates
for physics beyond the SM.

However, there are have several unattractive points in SUSY GUTs.
It should be noticed that 
the typical energy scale where the three
SM gauge couplings are unified is around $10^{16}$ GeV in ordinary SUSY
GUTs.  
Due to the decoupling theorem, the effects of the
GUT-scale particles are negligible at the TeV scale \cite{Appelquist:1974tg}.
Remnants of physics realized at the GUT scale can be investigated only
through relations among the masses and couplings 
of TeV-scale particles.  
Although 
the colored Higgs triplets and the electroweak Higgs doublets
originate from the same multiplets,
an unnaturally huge mass splitting between them
are supposed for suppressing proton decay.
To address this this doublet-triplet splitting problem
many ideas have been proposed
\cite{DW,SlidingSinglet,missingPARTN,pNG,orbifoldGUTs,Kakizaki:2001en}.

On the contrary, 
in the model where the doublet-triplet mass splitting is
realized by supersymmetrizing the Grand Gauge-Higgs
Unification (GHU) model,
the existence of new light
particles whose masses are of the order of the TeV scale is predicted
\cite{gGHU-DTS}.  
The Grand Gauge-Higgs Unification is constructed on an extra dimension
whose compactification scale is around the GUT scale \cite{gGHU},
and the GUT gauge symmetry is broken due to
the Hosotani mechanism \cite{hosotani}:
The non-trivial vacuum expectation value (VEV) of the extra-dimensional
component
of one of the gauge fields accounts for the GUT symmetry breaking.

In the Supersymmetric Grand Gauge-Higgs Unification (SGGHU) model, 
there appear a color octet chiral superfield, an $SU(2)_L$ triplet
chiral superfield with hypercharge zero and a neutral singlet chiral superfield
at the TeV scale as a by-product.  
As compared to the Minimal
Supersymmetric Standard Model (MSSM), 
the Higgs sector is extended with
the triplet and singlet chiral superfields.

In this presentation, we focus on the properties of the Higgs sector
of the SGGHU model, and
discuss its phenomenological signatures expected at collider
experiments.  
The masses and couplings of the SGGHU Higgs sector
particles are determined 
by solving renormalization group equations (RGEs) from the GUT
scale to the electroweak scale.  
We emphasize that models beyond the SM
can be distinguished by precisely
measuring the masses and couplings of the Higgs bosons at the
LHC and future electron-positron colliders such as the International
Linear Collider (ILC) \cite{ILC} and the CLIC \cite{CLIC}.
The SGGHU　model is a good example to show that collider experiments
are capable of testing GUT-scale physics.

\begin{figure}[t]
\centering
\includegraphics[width=70mm]{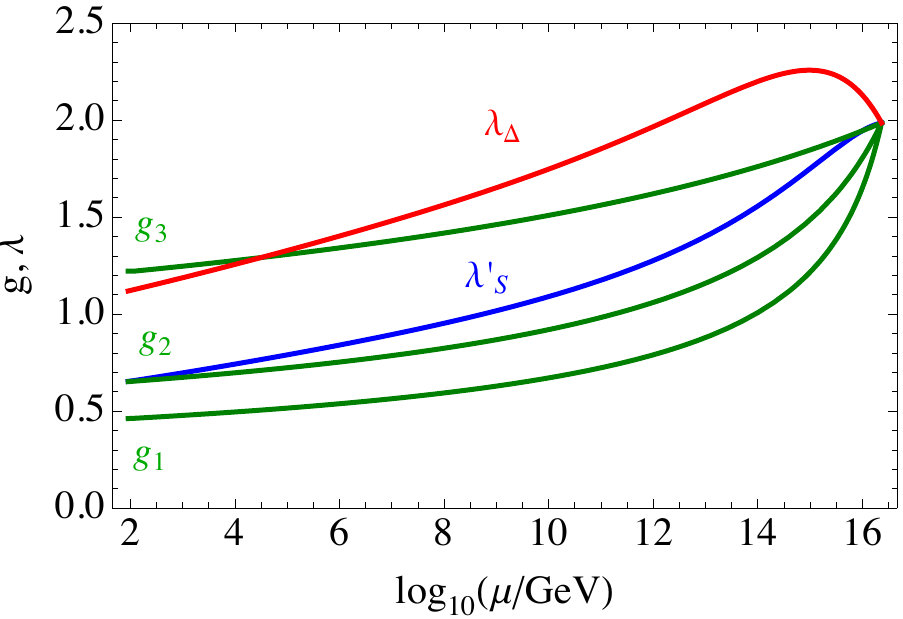}
\caption{\footnotesize RG evolution of the Higgs triplet and
  singlet couplings $\lambda_\Delta^{}$ (red line) and
  $\lambda'_S$ (blue) as well as the gauge couplings $g_3$,
  $g_2$ and $g_1$ (green) at the one loop level 
  as a function of the energy scale $\mu$.}
\label{fig:couplings}
\end{figure}

\section{Supersymmetric Grand Gauge-Higgs Unification}
\label{Sec:Model}

Let us discuss the Higgs sector of the low
energy effective theory of the SGGHU model.
At the TeV scale, the SGGHU Higgs sector consists of 
an $SU(2)_L$
triplet chiral superfield $\Delta$ and an neutral singlet chiral
superfield $S$ as well as the two MSSM Higgs doublets $H_u$ and $H_d$.
The superpotential is given by
\begin{eqnarray} \label{eq:WHiggs}
  W&=&\mu H_u \cdot H_d+\mu_\Delta^{}{\rm tr}(\Delta^2)+\frac{\mu_S^{}}2S^2
  \nonumber \\ 
  && +\lambda_\Delta^{} H_u \cdot \Delta H_d + \lambda_S^{}SH_u \cdot H_d \, ,
\end{eqnarray}
where $\Delta = \Delta^a \sigma^a/2$ with $\sigma^a$ $(a=1,2,3)$ being the
Pauli matrices.  
Since the triplet $\Delta$ and singlet $S$ originate from the gauge
supermultiplet, the following remarkable features are predicted.
Trilinear self-couplings among $S$ and $\Delta$ vanish although
they are not prohibited in
the general Higgs superpotential with the triplet and singlet
superfields.
The Higgs couplings $\lambda_\Delta^{}$ and $\lambda_S^{}$ are unified
with the SM gauge couplings at the GUT scale.  
Therefore,
the properties of the Higgs bosons are predicted with less
ambiguity.  The soft SUSY breaking operators
 in the Higgs potential are given by
\begin{eqnarray} \label{eq:VsoftHiggs}
  V_{\rm soft} 
  &=& \widetilde{m}_{H_d}^2 |H_d|^2 +\widetilde{m}_{H_u}^2|H_u|^2
  \nonumber \\
  &&+ 2 \widetilde{m}_\Delta^2 {\rm tr} (\Delta^\dag \Delta)
  +\widetilde{m}_S^2|S|^2 \nonumber \\
  &&+\left[B\mu H_u\cdot H_d +\xi S 
    + B_\Delta^{} \mu_\Delta^{}{\rm tr}(\Delta^2)  \right.
  \nonumber \\
  && +\frac{1}{2}B_S^{} \mu_S^{} S^2 
  +\lambda_\Delta^{}A_\Delta^{} H_u\cdot \Delta H_d 
  \nonumber \\
  && \left. +\lambda_S^{} A_S^{} S H_u\cdot H_d +{\rm h.c.} \right].
\end{eqnarray}
The soft parameters at the TeV scale are computed
by solving the RGEs.  
After radiative electroweak symmetry breaking,
four CP-even, three CP-odd and three charged Higgs bosons
appear as physical particles.  
The VEV of the neutral component of the
triplet Higgs boson $v_\Delta^{}$, which is derived from the minimization
conditions of the Higgs potential, must be less than $\simeq
10~{\rm GeV}$ in order to be consistent with the rho parameter constraint.  
Since $v_\Delta$ is sufficiently small compared to $v=246~{\rm GeV}$, 
it is negligible in the computations of the Higgs boson masses and
couplings.

\section{Analysis of  the Renormalization Group Equations}
\label{Sec:RGEAnalysis}

Here, we discuss RG evolution of couplings
and masses in the SGGHU model.  
The introduction of the light adjoint multiplets disturbs the successful gauge
coupling unification.  
One can easily recover the gauge coupling unification
by adding extra
incomplete $SU(5)$ matter multiplets.
A judicious choice for the matter
multiplets is a set of two vectorlike pairs of $(\bar{L},L)$
$(({\bf1},{\bf2})_{-1/2})$, one of $(\bar{U},U)$ $((\bar
{\bf3},{\bf1})_{-2/3})$ and one of $(\bar{E},E)$
$(({\bf1},{\bf1})_{1})$. Here, the numbers in the parentheses denote
$SU(3)_C$, $SU(2)_L$ and $U(1)_Y$ quantum numbers, respectively
\cite{gGHU-DTS}.  
Fig.~\ref{fig:couplings} shows the RG evolution of the
Higgs triplet and singlet couplings $\lambda_\Delta^{}$ (red
line) and $\lambda_S^{}$ (blue) as well as the gauge couplings
$g_3^{}$ , $g_2^{}$ and $g_1^{}$ (green) at the one loop
level as a function of the energy scale $\mu$.  The couplings are
normalized such that $\lambda'_S=(2\sqrt{5/3}) \lambda_S^{}$ for the
singlet Higgs coupling and $g_1=(\sqrt{5/3})g_Y$ for the $U(1)_Y$
gauge coupling.  The resulting Higgs trilinear couplings at the
TeV-scale are given by
\begin{equation} \label{eq:couplings}
  \lambda_\Delta^{}=1.1\, , \qquad \lambda_S^{}=0.25\, .
\end{equation}

The soft SUSY breaking parameters at the TeV scale are also derived
by solving the RGEs.  
As shown in Fig.~\ref{fig:couplings},
the gauge couplings become strong around the GUT scale.
Since the unified gaugino mass at the GUT scale has to be large
in order to avoid the gluino mass bound \cite{LHCgluino}, 
typical values for the 
soft sfermion and Higgs masses at
the SUSY breaking scale are in the multi-TeV range.  
Therefore, 
some tuning is needed for successful radiative electroweak symmetry
breaking.  
In spite of such difficulties, we can also
obtain soft Higgs mass parameters of the order of ${\cal O}(100)$ GeV by tuning
among the input parameters at the GUT scale, as shown in the next section.

\section{Impact on the Properties of the Higgs Sector}

The prediction about the SM-like Higgs boson mass is affected by its
interactions with the $SU(2)_L$ triplet and singlet Higgs multiplets.
When the soft scalar masses of the triplet and singlet Higgs multiplets
are relatively large, the approximate formula for 
the SM-like Higgs boson mass is
\cite{MSSMHiggsMass,Espinosa}
\begin{eqnarray} \label{eq:higgs_mass}
  m_h^2 &\simeq&
m_Z^2 \cos^2 \beta 
\nonumber \\
&& +\frac{3 m_t^4}{2\pi^2 v^2}\left( \ln{\frac{m_{\tilde{t}}^2}{m_t^2}} +
  \frac{X_t^2}{m_{\tilde{t}}^2} \left(1 -\frac{X_t^2}{12m_{\tilde{t}}^2} 
    \right) \right)
\nonumber \\
&& +\frac{1}{8}\lambda_\Delta^2 v^2 \sin^2 {2\beta}
+\frac{1}{2}\lambda_S^2 v^2 \sin^2 {2\beta}
\, ,
\end{eqnarray}
where $m_Z$ is the $Z$-boson mass, $m_t$ is the top quark mass,
$m_{\tilde{t}}$ is the geometrical average of the stop mass eigenvalues, 
and $X_t = A_t - \mu \cot \beta$.  
In the MSSM, large stop masses are required even in the maximal stop
mixing case in order to obtain an SM-like Higgs boson mass of 125 GeV
\cite{MSSMHiggsMass2}.  In our model, on the contrary, the SM-like
Higgs boson mass is lifted up by the Higgs trilinear interactions with
the triplet and singlet superfields for small $\tan \beta$.  Notice
that the same mechanism is realized in the next-to-MSSM (NMSSM)
\cite{NMSSM}.  For our numerical computations of the masses of the
Higgs bosons and their superpartners, we have appropriately modified
the public numerical code \texttt{SuSpect} \cite{suspect} by adding
the contributions from the triplet and singlet Higgs superfields.
Since some fine tuning for the GUT-scale input parameters
is needed, 
we show our numerical results based on several benchmark points
that can reproduce the observed SM-like Higgs boson mass.
Because of theoretical uncertainties in the computation
of the SM-like Higgs boson mass,
we take $122~{\rm GeV} < m_h < 129~{\rm GeV}$
as its allowed mass range.
We focus on the following three typical cases:
\begin{itemize}
\item[(A)] Mixings between the SM-like Higgs boson and the other CP-even Higgs bosons are small.
\item[(B)] Mixings between the SM-like Higgs bosons and the CP-even components of the triplet and singlet Higgs fields are small.
\item[(C)] The CP-even components of the triplet and singlet Higgs fields affect the SM-like Higgs boson couplings.
\end{itemize}
Three successful benchmark points for the GUT-scale input parameters and the
TeV-scale parameters obtained after solving the RGEs are listed in
Tab. \ref{tab:benchmark-GUT} and \ref{tab:benchmark-TeV}, respectively.

\begin{table*}[t]
  \begin{flushleft}
{\footnotesize
  \begin{tabular}{|c||c|c|c|c|c|c|c|}
    \hline
    Case &
    $\tan \beta$ &
    $M_{1/2}$ &
    $\mu_{\Sigma}$
    \\ \hline  \hline
    (A)(B)(C) &
    $3$ &
    $3600~{\rm GeV}$ &
    $-300~{\rm GeV}$
    \\ \hline 
  \end{tabular}
  \begin{tabular}{|c||c|c|c|c|c|c|c|}
    \hline
    Case &
    $A_0$ & 
    $\widetilde{m}_0^2$ & 
    $\widetilde{m}_{H_u}^2$ &
    $\widetilde{m}_{H_d}^2$ &
    $\widetilde{m}_{5}^2$ &
    $\widetilde{m}_{10}^2$ &
    $\widetilde{m}_\Sigma^2$
    \\ \hline  \hline
    (A) &
    $5500~{\rm GeV}$ & 
    $(1000~{\rm GeV})^2$ &
    $(10375~{\rm GeV})^2$ &
    $(8570~{\rm GeV})^2$ & 
    $- (6300~{\rm GeV})^2$ &
    $- (2000~{\rm GeV})^2$ &
    $- (570~{\rm GeV})^2$
    \\ \hline 
    (B) &
    $1000~{\rm GeV}$ & 
    $(1800~{\rm GeV})^2$ &
    $(12604~{\rm GeV})^2$ &
    $(10381.5~{\rm GeV})^2$ & 
    $- (7700~{\rm GeV})^2$ &
    $- (1960~{\rm GeV})^2$ &
    $- (670~{\rm GeV})^2$
    \\ \hline 
    (C) &
    $8000~{\rm GeV}$ & 
    $(3000~{\rm GeV})^2$ &
    $(10605.1~{\rm GeV})^2$ &
    $(8751.4~{\rm GeV})^2$ & 
    $- (6418~{\rm GeV})^2$ &
    $- (1638.5~{\rm GeV})^2$ &
    $- (400~{\rm GeV})^2$
        \\ \hline 
  \end{tabular}
}
\end{flushleft}
  \caption{\footnotesize Benchmark points for the input parameters at 
    the GUT scale.}
  \label{tab:benchmark-GUT}
\end{table*}

\begin{table*}[t]
  \begin{flushleft}
    {\footnotesize
  \begin{tabular}{|c||c|c|c|c|c|c|c|c|}
    \hline
    Case &
    $M_1$ &
    $M_2$ &
    $M_3$ &
    $\mu_{\Delta}$ &
    $\mu_S^{}$
    \\ \hline  \hline
    (A)(B)(C) &
    $194~{\rm GeV}$ & 
    $388~{\rm GeV}$ &
    $1360~{\rm GeV}$ &
    $-252~{\rm GeV}$ &
    $-85.8~{\rm GeV}$
    \\ \hline 
\end{tabular}
  \begin{tabular}{|c||c|c|c|c|c|c|c|c|}
    \hline
    Case &
    $\mu$ &
    $B \mu$ & 
    $\widetilde{m}_{u_3}$ & 
    $\widetilde{m}_{q_3}$ &
    $y_t A_t$
    \\ \hline  \hline
    (A) &
    $205~{\rm GeV}$ &
    $41400~{\rm GeV}^2$ &
    $3290~{\rm GeV}$ & 
    $4830~{\rm GeV}$ &
    $4030~{\rm GeV}$
    \\ \hline 
    (B) &
    $177~{\rm GeV}$ &
    $40800~{\rm GeV}^2$ &
    $1730~{\rm GeV}$ & 
    $4480~{\rm GeV}$ &
    $6050~{\rm GeV}$
    \\ \hline 
    (C) &
    $174~{\rm GeV}$ &
    $42000~{\rm GeV}^2$ &
    $4220~{\rm GeV}$ & 
    $5550~{\rm GeV}$ &
    $2910~{\rm GeV}$
    \\ \hline 
  \end{tabular}
  \begin{tabular}{|c||c|c|c|c|c|c||c|}
    \hline
    Case &
    $\widetilde{m}_{\Delta}$ & 
    $\widetilde{m}_{S}$ &
    $\lambda_\Delta^{} A_\Delta^{}$ &
    $\lambda'_S A_S^{}$ &
    $B_\Delta^{} \mu_\Delta^{}$ &
    $B_S^{} \mu_S^{}$ &
    $m_h$
    \\ \hline  \hline
    (A) &
    $607~{\rm GeV}$ & 
    $805~{\rm GeV}$ &
    $662~{\rm GeV}$ &
    $683~{\rm GeV}$ &
    $92000~{\rm GeV}^2$ &
    $-78700~{\rm GeV}^2$ &
    $123~{\rm GeV}$
    \\ \hline 
    (B) &
    $784~{\rm GeV}$ & 
    $612~{\rm GeV}$ &
    $1340~{\rm GeV}$ &
    $1110~{\rm GeV}$ &
    $30700~{\rm GeV}^2$ &
    $-110000~{\rm GeV}^2$ &
    $123~{\rm GeV}$
    \\ \hline 
    (C) &
    $521~{\rm GeV}$ & 
    $216~{\rm GeV}$ &
    $284~{\rm GeV}$ &
    $446~{\rm GeV}$ &
    $207000~{\rm GeV}^2$ &
    $-33600~{\rm GeV}^2$ &
    $122~{\rm GeV}$
    \\ \hline 
  \end{tabular}
}
  \end{flushleft}
  \caption{\footnotesize TeV-scale parameters obtained by solving the RGEs.}
  \label{tab:benchmark-TeV}
\end{table*}

\begin{figure*}[t!]
  \begin{center}
      \begin{minipage}{0.45\hsize}
        \begin{center}
          \includegraphics[clip,width=75mm]{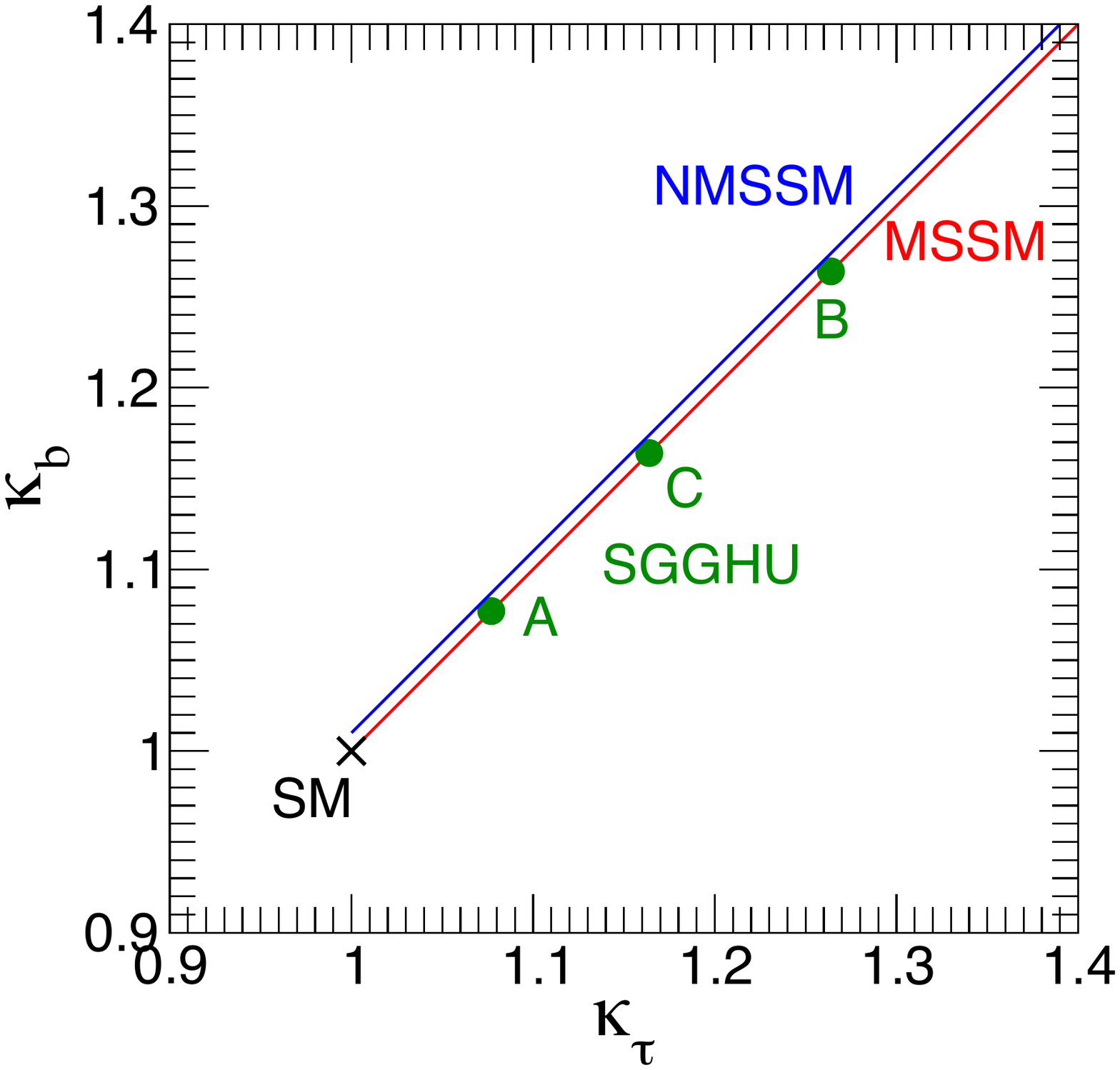}    
        \end{center}
      \end{minipage}
      \begin{minipage}{0.45\hsize}
        \begin{center}
          \includegraphics[clip,width=75mm]{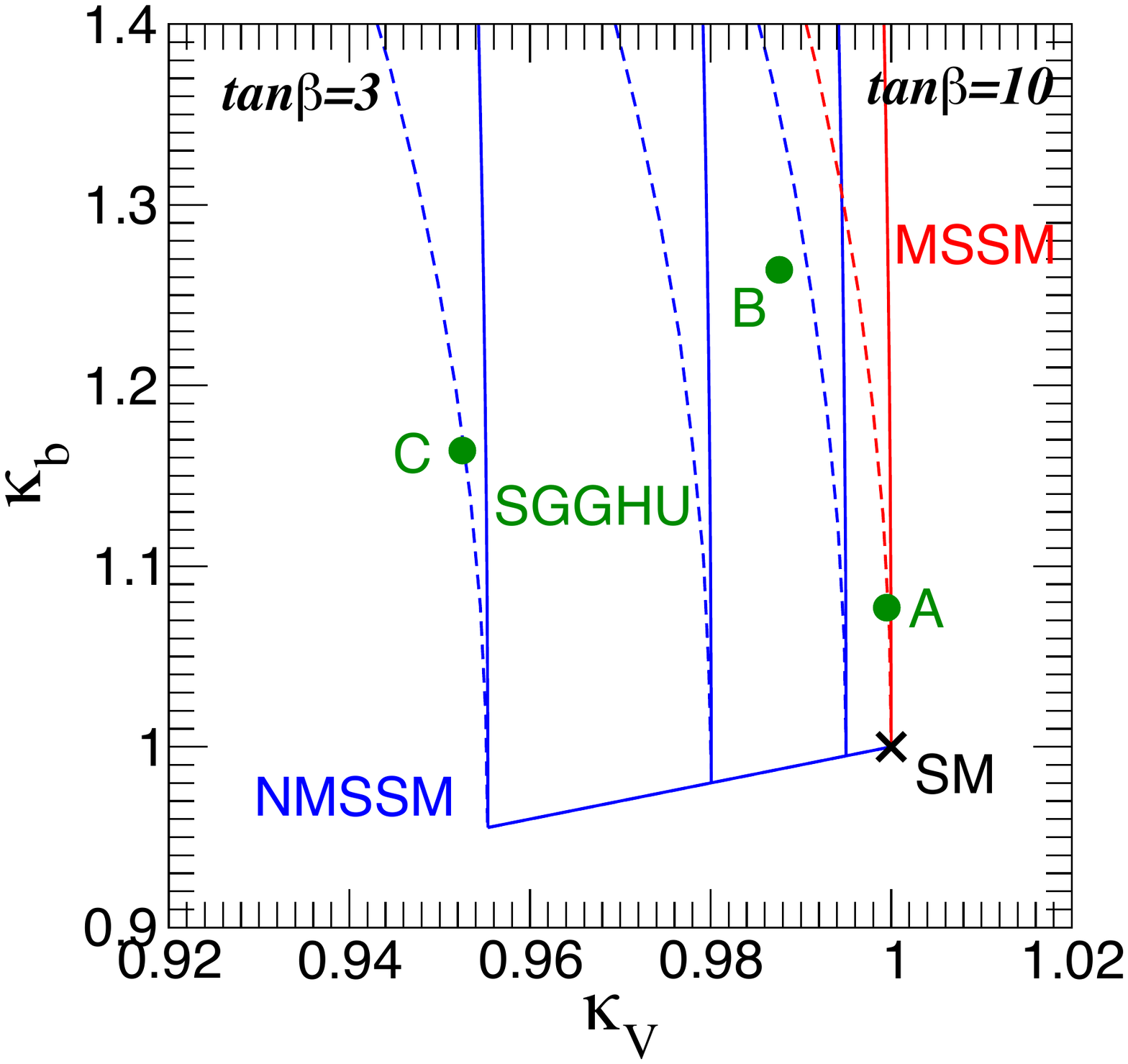}    
        \end{center}
      \end{minipage}
      \begin{minipage}{0.45\hsize}
        \begin{center}
          \includegraphics[clip,width=75mm]{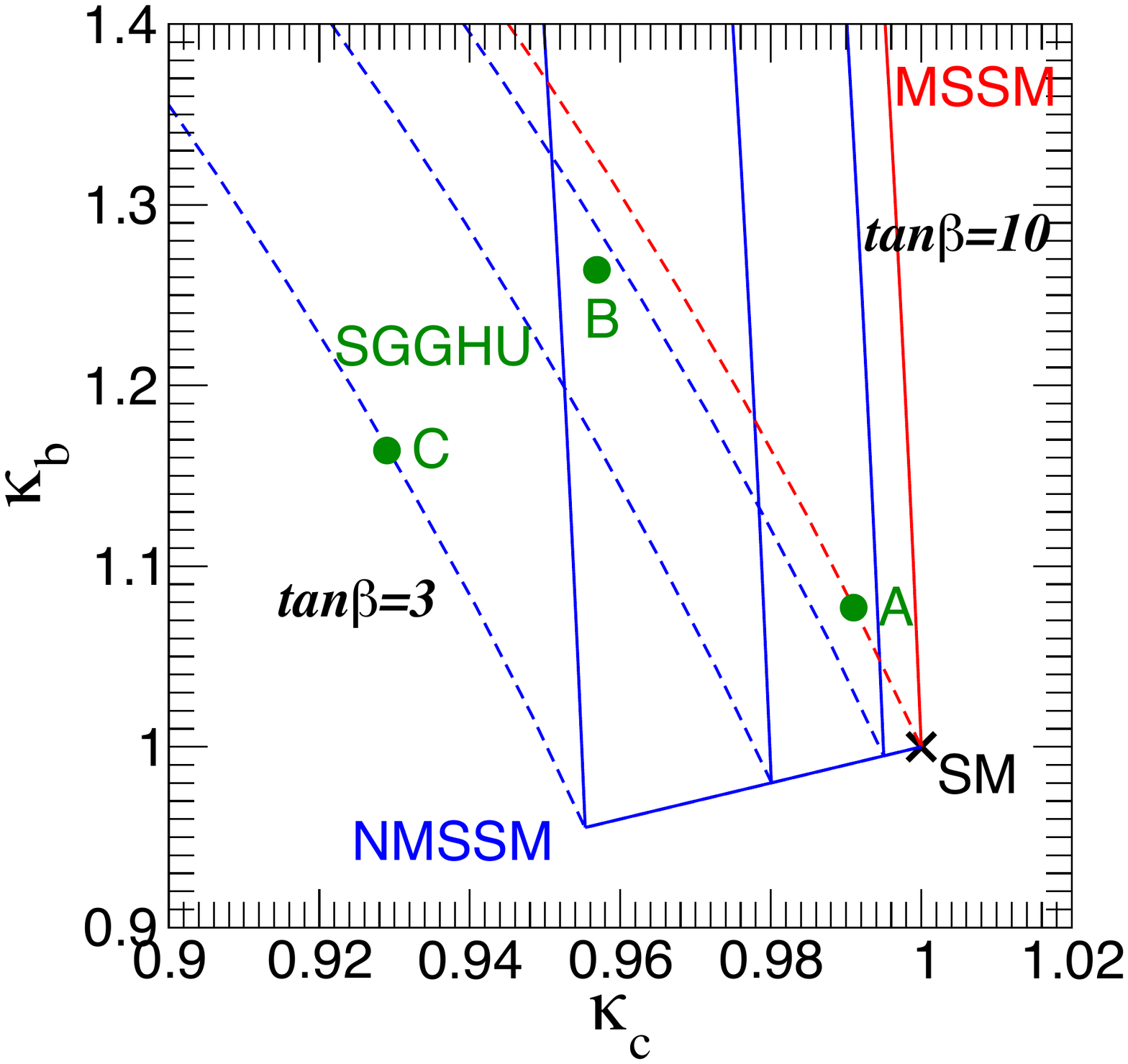}    
        \end{center}
      \end{minipage}

      \caption{\footnotesize The scaling factors
        $\kappa_X^{}$ are plotted on the
        $\kappa_\tau^{}$-$\kappa_b^{}$ plane, the
        $\kappa_V^{}$-$\kappa_b^{}$ plane $(V=Z,W)$, and the
        $\kappa_c^{}$-$\kappa_b^{}$ plane.  The deviations in the
        three benchmark scenarios (A), (B) and (C) in the SGGHU are
        shown with green blobs.  The MSSM predictions are shown with
        red lines for $\tan \beta=10$ (thick line) and $\tan \beta=3$
        (dashed).  The NMSSM predictions are shown with blue grid
        lines for $\tan \beta=10$ (thick) and $\tan \beta=3$ (dashed),
        which indicate mixings between the SM-like and singlet-like
        Higgs bosons of 10\%, 20\% and 30\% from the right to the
        left. }
    \label{fig:fingerprint}
  \end{center}
\end{figure*}

The couplings between the SM-like Higgs boson and SM particles can be
significantly altered due to the existence of the triplet
and singlet Higgs bosons.  
In discussing the SM-like Higgs boson couplings, it is useful to
introduce the scaling factors defined as
\begin{eqnarray}
  \kappa_X^{} = \frac{g_{hXX}^{}}{g_{hXX}^{}|_{\rm SM}^{}} \, ,
\end{eqnarray}
where $g_{hXX}^{}$ is the coupling with the SM particle $X$.  
Fig.~\ref{fig:fingerprint} shows the deviations in the scaling factors
$\kappa_X^{}$ are plotted on the $\kappa_\tau^{}$-$\kappa_b^{}$ plane,
the $\kappa_V^{}$-$\kappa_b^{}$ plane $(V=Z,W)$ and the
$\kappa_c^{}$-$\kappa_b^{}$ plane.
The deviations in the three SGGHU benchmark points (A), (B)
and (C) are shown with green blobs.  
The MSSM predictions
are displayed with red lines for $\tan \beta=10$ (thick line) and $\tan
\beta=3$ (dashed).  
The NMSSM predictions are shown with blue grid
lines for $\tan \beta=10$ (thick) and $\tan \beta=3$ (dashed), which
denote mixings between the SM-like and singlet-like Higgs bosons of
10\%, 20\% and 30\% from the right to the left.  
Notice that the SM-like Higgs boson
couplings to the down-type quarks and charged leptons are common in
this model, as in the type-II two-Higgs-doublet model.
Therefore, the MSSM, NMSSM and SGGHU predict 
$\kappa_b^{}/\kappa_\tau^{}=1$ at the tree level.
At the ILC with
$\sqrt{s}=500~{\rm GeV}$, expected accuracies for the Higgs boson couplings
are 
$\kappa_Z$, $\kappa_W$, $\kappa_b^{}$, $\kappa_\tau^{}$ and
$\kappa_c^{}$ are 1.0\% 1.1\% 1.6\%, 2.3\% and 2.8\%, respectively
\cite{ILCHiggs}.  
Fig.~\ref{fig:fingerprint} shows that typical SGGHU
predictions about the scaling factors can be distinguished
from the corresponding SM and MSSM predictions 
through precision measurements of the Higgs boson couplings
at the future ILC.  
It may be difficult to completely distinguish
our model from the NMSSM only from the Higgs boson couplings.
However, if the pattern of the deviations of the Higgs couplings is found
to be close to one of our benchmark scenarios, the possibility of the
SGGHU is increased.  
Independent measurement of $\tan \beta$ using
Higgs boson decay at the ILC \cite{Gunion:2002ip,Kanemura:2013eja}
will be also helpful in discriminating TeV-scale models.  
For the above benchmark points, the predicted range of the 
Higgs boson coupling with the photon and that of the Higgs self-coupling are
$0.94 < \kappa_\gamma^{} < 1.0$ and $0.82 < \kappa_h^{} < 0.93$,
respectively.
More precise measurements at the ILC with
$\sqrt{s}=1~{\rm TeV}$ \cite{ILCHiggs} are needed in order to 
observe such small deviations.

\begin{figure}[t]
\includegraphics[width=80mm]{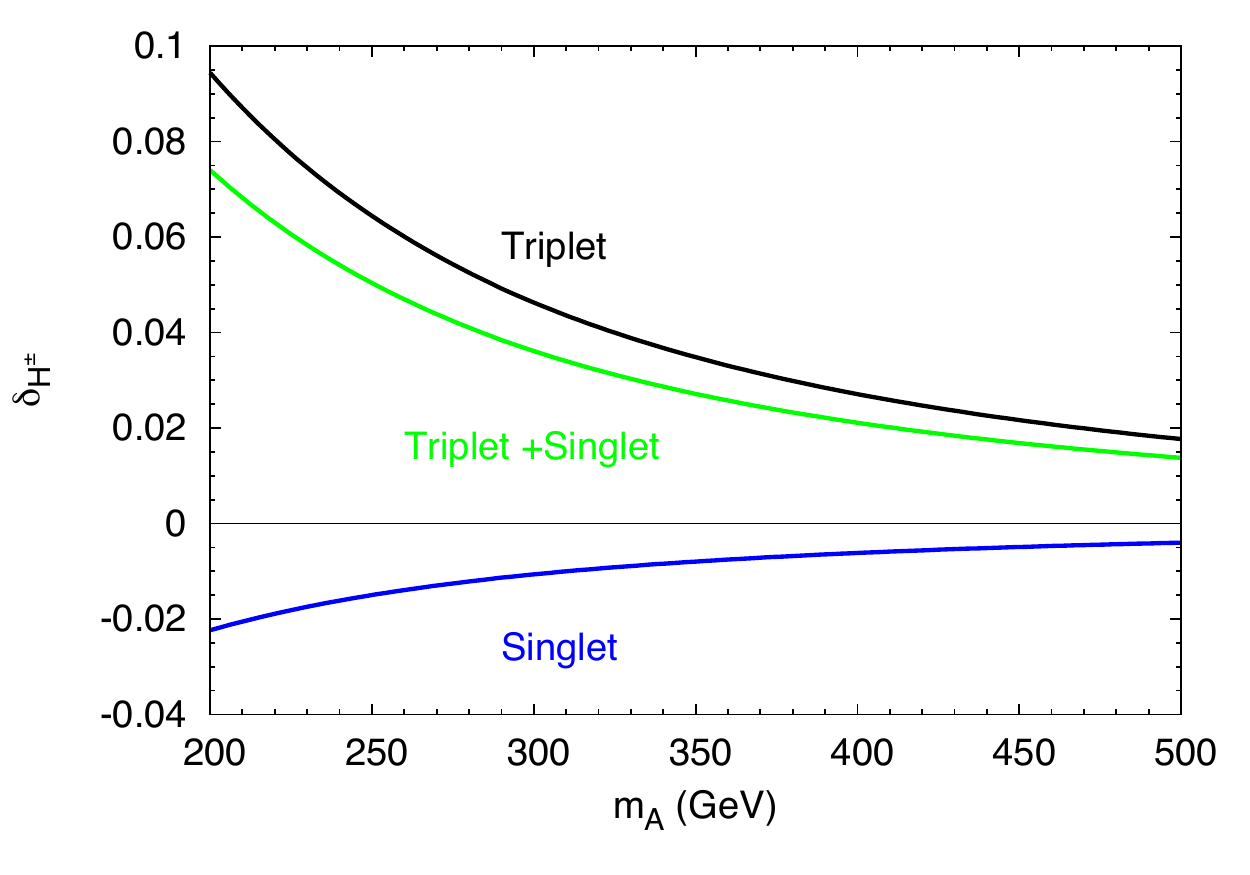}
\caption{\footnotesize The mass deviation parameter $\delta_{H^\pm}^{}$
  as a function of the
  MSSM-like CP-odd Higgs boson mass $m_A^{}$ for relatively large
  soft Higgs masses.  The black, blue and green lines represent triplet
  contribution, singlet contribution and their sum, respectively.}
\label{fig:dhpm_ma}
\end{figure}

Let us turn to discussions on the masses of the additional MSSM-like Higgs
bosons.  
Given relatively large soft scalar masses of the Higgs triplet and
singlet fields, the approximate formula for the MSSM-like charged
Higgs boson mass $m_{H^\pm}^{}$ is given by
\begin{eqnarray} 
  m_{H^\pm}^2
  & = & m_{H^\pm}^2|_{\rm MSSM}^{} (1 +\delta_{H^\pm}^{})^2  
  \nonumber \\
  &\simeq & m_A^2 +m_W^2 +\frac{1}{8}\lambda_\Delta^2 v^2
  -\frac{1}{2}\lambda_S^2v^2\, ,
\end{eqnarray}
where $m_A$ stands for the MSSM-like CP-odd Higgs boson mass, and
$\delta_{H^\pm}^{}$ parametrizes the deviation of $m_{H^\pm}^{}$ from
its MSSM prediction.  
The sign difference between the triplet and
singlet contributions arises from group theoretical factors.  
Since we have
$\lambda_\Delta >\lambda_S/2$ due to
radiative corrections, the MSSM-like charged Higgs boson mass in our
model is larger than its MSSM prediction.  
Fig.~\ref{fig:dhpm_ma}
shows the mass deviation parameter $\delta_{H^\pm}^{}$ 
as a function of $m_A^{}$ for relatively large soft
Higgs masses.  
The black, blue and green lines represent triplet
contribution, singlet contribution and their sum, respectively.
When the masses of the MSSM-like Higgs bosons are smaller than
$500~{\rm GeV}$, the deviation parameter is $\delta_{H^\pm}^{} \sim O(1)~\%$ -
$O(10)~\%$ and measurable at the LHC \cite{HiggsWG}.

If the masses of the triplet-like and singlet-like Higgs bosons are smaller than
$500~{\rm GeV}$, these new particles can be directly produced at the
ILC and CLIC.  As shown in Tab.~\ref{tb:spectrum}, 
such light Higgs bosons are realized in
the benchmark scenario (C).  For instance, $\Delta^{\pm}$ can
be probed through the channel $e^+ e^- \to \Delta^+ \Delta^- \to
t\bar{b} \bar{t} b$, which is induced by the mixing between the
triplet-like and MSSM-like charged Higgs bosons.

\begin{table}[t]
  \begin{center}
  \begin{tabular}{|c|c|c|}
    \hline
    CP-even & CP-odd & Charged \\ \hline \hline
    $122~{\rm GeV}$     & $-$                & $-$         \\ \hline
    $139~{\rm GeV}$     & $171~{\rm GeV}$    & $204~{\rm GeV}$     \\ \hline
    $370~{\rm GeV}$     & $304~{\rm GeV}$    & $496~{\rm GeV}$     \\ \hline
    $745~{\rm GeV}$     & $497~{\rm GeV}$    & $745~{\rm GeV}$     \\ 
    \hline
  \end{tabular}
  \caption{The mass spectrum of the Higgs bosons for the benchmark scenario (C).} 
  \label{tb:spectrum}
      \end{center}
\end{table}

As discussed above, 
comprehensive analysis of the masses and couplings of the Higgs
bosons at the LHC and future electron-positron colliders can
distinguish models realized at the TeV scale.
Even if the additional Higgs bosons are beyond the reach of direct
discovery, their effects are left 
in the SM-like Higgs boson couplings and the MSSM Higgs masses
and can be indirectly probed by their precision measurements.
Therefore, a new electron-positron collider is mandatory
for exploring the Higgs sector and the underlying theory.

\section{Summary}

In the SUSY $SU(5)$ GUT model where the Hosotani mechanism accounts 
for the GUT symmetry breaking, the low-energy Higgs sector
involves a Higgs triplet and singlet chiral
superfields as well as the two MSSM Higgs doublets.  
We have evaluated
the SM-like Higgs boson couplings to SM particles.  
It is shown that these couplings deviate from the corresponding SM
predictions by ${\cal O}(1)\%$ when the triplet and
singlet Higgs bosons are lighter than $\simeq 1~{\rm TeV}$.
Such deviations are measurable at future electron-positron colliders.
When the masses of the MSSM-like charged Higgs boson and the MSSM-like
CP-odd Higgs boson are less than $\simeq 500~{\rm GeV}$, their mass
difference is larger than the MSSM prediction by ${\cal O}(1)\%$ -
${\cal O}(10)\%$, and can be measured at the LHC.  Using these
observations, our model can be distinguished from the MSSM and NMSSM.
We emphasise that our supersymmetric grand gauge-Higgs unification
model is a good example to show that colliders are able to test
physics realized at the GUT scale.

\section*{Acknowledgments}
The author would like to thank Shinya Kanemura, Hiroyuki Taniguchi and
Toshifumi Yamashita for the fruitful collaboration on this project.
The author was supported in part by Grant-in-Aid for Scientific
Research from Ministry of Education, Culture, Sports, Science and
Technology (MEXT), Japan, No. 26104702.







\end{document}